\documentclass[twocolumn,showpacs,preprintnumbers,amsmath,amssymb]{revtex4}
\usepackage{bm}
\usepackage[dvips]{graphicx}
\usepackage{epsfig}
\begin{document}
\title{Extended states in a one-dimensional generalized dimer model}
\author{P. Ojeda, R. Huerta-Quintanilla and M. Rodr\'{\i}guez-Achach}
\affiliation{\it Departamento de F\'{\i}sica Aplicada Unidad M\'erida, 
Cinvestav-IPN,
Km.\ 6 Carr.\ Antigua a Progreso, M\'erida, Yucat\'an, M\'exico.}
\begin{abstract}
The transmission coefficient for one dimensional systems is given
in terms of Chebyshev polynomials using the tight binding model.
This result is applied to a system composed of two impurities
located between $N$ sites of a host lattice. It is found that the system
has extended states for several values of the energy. Analytical expressions
are given for the impurity site energy in terms of the electron's energy.
The number of resonant states grow like the number of host sites between
the impurities. This property makes the system interesting since it is 
a simple task to design a configuration with resonant energy very close
to the Fermi level $E_F$.  
\end{abstract}


\maketitle


Since the seminal paper of Anderson \cite{anderson}, the problem
of localization has been fundamental for the understanding of 
electronic transport in one dimensional disordered systems. 
The discovery of extended or delocalized states in systems 
with correlated disorder has renewed the interest in the 
transport properties of electrons  for particular 
configurations of the lattice. Different types of systems 
[2-10]
have been demonstrated to have extended states either with
diagonal, or non-diagonal disorder or both.

One of the first systems studied was the so-called random
dimer model (RDM) \cite{flores,dunlap}, in which a certain number
of dimers (two adjacent impurities) were introduced randomly 
into the host lattice. Extended states with resonant energies 
equal to the site energy of the impurity were found.

In this paper, we describe a system consisting of two impurities
hosted in a regular lattice and separated by $N$ lattice
sites. This system is studied using the tight-binding
Hamiltonian and the transfer matrix method. In addition, we use
the Cayley-Hamilton theorem to express the transmission
coefficient in terms of the Chebyshev polynomials
of the second kind \cite{arfken}. The transmission coefficient can be related
to the conductance of the system through the Landauer
formula \cite{landauer}.


Consider a general $n\times n$ matrix $M$ whose characteristic
equation can always be written as
\begin{equation}
\lambda^n = C_{n-1}\lambda^{n-1}+C_{n-2}\lambda^{n-2}+\cdots C_0.
\end{equation}
We can write
\begin{eqnarray}
\nonumber
\lambda^{n+1}&=&\lambda^n\lambda\\
\nonumber
&=&(C_{n-1}^2+C_{n-2})\lambda^{n-1}\\
&&+(C_{n-1}C_{n-2}+C_{n-3})\lambda^{n-2}
   +\cdots
\end{eqnarray}
and so on for higher powers of $\lambda$. In general the above equation
will have the form
\begin{equation}
\lambda^{n+m}=p_{n+m-1}\lambda^{n-1}+p_{n+m-2}\lambda^{n-2}+\cdots
              +p_m\lambda^0,
\end{equation}
where the $p_j$ are polynomials in the trace and the minors
of $M$. Now recall the Cayley-Hamilton theorem \cite{ma}: if $p(\lambda)=0$ is
the characteristic equation of $M$, then $p(M)=0$, that is, the matrix $M$
is also a ``root" of the characteristic equation. The implication
of this is that we can always write powers $M^{n+N}$ in terms of simpler
expressions involving $M^{n-1}$ and the polynomials $p_j$.

In the case of a rank 2 transfer matrix
with determinant equal to one, the characteristic equation is
\begin{equation}
\lambda^2=2x\lambda-1,
\end{equation}
where $2x = \mbox{Tr} M$. It is easy to verify that higher powers of
$\lambda$ are
\begin{eqnarray}
\lambda^3&=&(4x^2-1)\lambda - 2x,\\
\lambda^4&=&(8x^3-4x)\lambda - (4x^2-1),\\
\vdots&&\qquad\vdots\\
\lambda^n&=&p_{n-1}\lambda -p_{n-2},
\end{eqnarray}
and by virtue of the Cayley-Hamilton theorem, we can conclude
\begin{equation}
\label{cht}
M^n=p_{n-1}M -p_{n-2}.
\end{equation}
Note that the above formalism applies to a general matrix whose only
requirement is to have a unit determinant. In the case of rank two
matrices, the polynomials coincide with the second kind Chebyshev
polynomials.
The above results can be applied to either continuous \cite{pp1,sanchez} 
as well to discrete type lattices.

In the following we consider the discrete one dimensional chain in the
tight-binding approximation \cite{eco,marder}, for which we have the
Hamiltonian
\begin{equation}
H=\sum_n\epsilon_n|n\rangle\langle n|+V\sum_n\left[|n\rangle\langle n+1|+
|n\rangle\langle n-1|\right]
\end{equation}
where $\epsilon$ is the site energy and $V$ is the hopping amplitude
between nearest sites. The resultant Schr\"{o}dinger equation is
\begin{equation}
i\dot c_n(t)=\epsilon_nc_n(t)+c_{n+1}(t)+c_{n-1}(t),
\end{equation}
where $c_n(t)$ is the probability amplitude for an electron to
be at site $n$. After the
substitution $c_n(t)=C_n\exp(-iEt)$, where $E$ is the energy of the
electron, we get the equation
\begin{equation}
\label{recrel}
C_{n+1}+C_{n-1}=\left(\frac{E-\epsilon_n}{V}\right)C_n,
\end{equation}
with the equivalent matrix representation
\begin{equation}
\label{trans}
\left(\begin{array}{c}
C_{n+1}\\
C_n\end{array}\right)=\left(\begin{array}{cc}
                      2x_n & -1\\
                      1       &  0 \end{array}\right)
\left(\begin{array}{c}
C_{n}\\
C_{n-1}\end{array}\right),
\end{equation}
where $x_n=\frac{E-\epsilon_n}{2V}$. The above matrix is called the 
transfer matrix $M$ of the site $n$. If we define
\begin{equation}
X_{n+1}=\left(\begin{array}{c}
C_{n+1}\\
C_n
\end{array}\right),
\end{equation}
equation (\ref{recrel}) takes the more compact form
\begin{equation}
\label{compact}
X_{n+1}=M_nX_n,
\end{equation}
and we will have
\begin{equation}
\label{total}
X_{N+1}=\left(\prod_{i=1}^{N}M_i\right)X_1.
\end{equation}
Consider an electron impinging an a sample that begins at site $n=1$ 
and ends at site $n=N$. The electron amplitudes will be given by
\begin{equation}
c_n=\left\{\begin{array}{ccc}
Ae^{ikn}+Be^{-ikn}&;&n\le 1\\
Ce^{ikn}+De^{-ikn}&;&n\ge N
\end{array}\right.
\end{equation}
The transmission coefficient $T$ will be given by $|C/A|^2$
and we obtain
\begin{eqnarray}
T(N,x)&=&4(1-x^2)(\left[(m_{21}-m_{12})+(m_{22}-m_{11})x\right]^2 \nonumber\\
\label{trc}
          &&+(m_{11}+m_{22})^2(1-x^2))^{-1},
\end{eqnarray}
where the $m_{ij}$ are the elements of the total transfer matrix
given in (\ref{total}) and $x=E/2=\cos k$, so we are taking
$V=1$. Note that the $m_{ij}$ matrix elements contain all the
physical information about the system that we are considering as a 
sample.

\begin{figure}[t]
\begin{center}
\includegraphics[width=8cm]{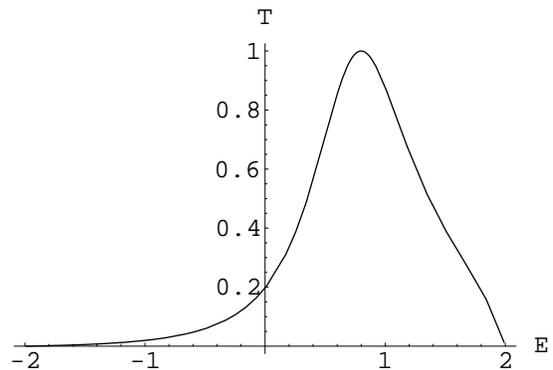}\\
\parbox{8.5cm}{\caption{ \label{fig1}
\footnotesize {Transmission for the trimer with site energy $\epsilon_1=1.8.$
The resonance appears at $E=0.8$.}
}}
\end{center}
\end{figure}

If we consider a piece of a chain with $N$ sites of equal energy
$\epsilon_0$ embedded in a host lattice of site energy 0, 
sometimes referred in the literature as an
$N$-mer, equation (\ref{compact}) becomes 
\begin{equation}
X_{N+1}=M_0^NX_1,
\end{equation}
and we can apply the Cayley-Hamilton theorem to get
\begin{equation}
\label{ch}
M^N=\left(\begin{array}{cc}
2x_0U_{N-1}-U_{N-2} & -U_{N-1}\\
U_{N-1} & -U_{N-2}
\end{array}\right).
\end{equation}
where the Chebyshev polynomials depend on $x_0=(E-\epsilon_0)/2$.
Equation (\ref{trc}) becomes

\begin{eqnarray}
T(N,x,x_0)&=&(1-x^2)[U_{N-1}^2(1-xx_0)^2\nonumber\\
            &&+(x_0U_{N-1}-U_{N-2})^2(1-x^2)]^{-1}.
\end{eqnarray}
If we take the limit $\epsilon_0=0$, then $x=x_0$ and we
get a regular system (all site energies are the same). It is easy to see
that in this case the above equation gives $T=1$.
As an example of use of the above result, we consider a trimer with site
energy $\epsilon_0=1.8$. In Fig. (\ref{fig1}) we plot the transmission
as a function of the energy. We can see that $T=1$ at an energy of
0.8. These resonant values of energy are giving explicitly in
\cite{izrailev} as $E=\epsilon_0+2\cos\mu$, where $\mu=(i+1)\pi/N$ and
$i=0,1...N-2$. In our example we have $N=3$, therefore we have two resonances
at $E=0.8$ and $E=2.8$, the latter lies outside the band width.

Recently, there has been much attention devoted to the
RDM. In this model, two adjacent
impurities are put in an otherwise regular chain. The result is that
we have extended states present in the chain when the energy of these
states is equal to the site energy of the impurities. The
extended states are still there when the dimer is placed randomly
in the chain.

As another application of Eq.\ (\ref{trc}), we consider an example which consists in
introducing two impurities not necessarily adjacent in
a regular lattice. We call this system a ``generalized dimer".

In this system we have two impurities with equal site 
energies separated by a linear
chain of $N$ sites with energy equal to zero (The fact that the 
site energy is set to zero is not relevant 
since the site energy differences is what really matters).
Then  $x = E/2$ and $x_1=(E-\epsilon_1)/2$. The total
transfer matrix for the system will be of the form
\begin{equation}
M_T=M_1(M_0)^NM_1,
\end{equation}
where $M_1$ is the transfer matrix corresponding to the impurity
site and $M_0$ corresponds to the regular sites. We can now use
result (\ref{ch}) and obtain
\begin{equation}
\label{mtemp}
M_T=\left(\begin{array}{cc}
4x_1^2F_0-2x_1U_{N-1}-F1 & U_{N-1}-2x_1F_0\\
-U_{N-1}+2x_1F_0 & -F_0
\end{array}\right).
\end{equation}
where $x = E/2$, $x_1=(E-\epsilon_1)/2$, $F_0 = 2xU_{N-1}-U_{N-2}$
and $F_1 = 2x_1U_{N-1}-U_{N-2}$.
Now that we know the total transfer matrix elements, we can substitute
them in equation (\ref{trc}) in order to obtain the transmission
through the whole sample. It is clear that the polynomials are evaluated at
$x$ since they describe the regular part of the sample.

\begin{figure}[t]
\begin{center}
\includegraphics[width=8.7cm]{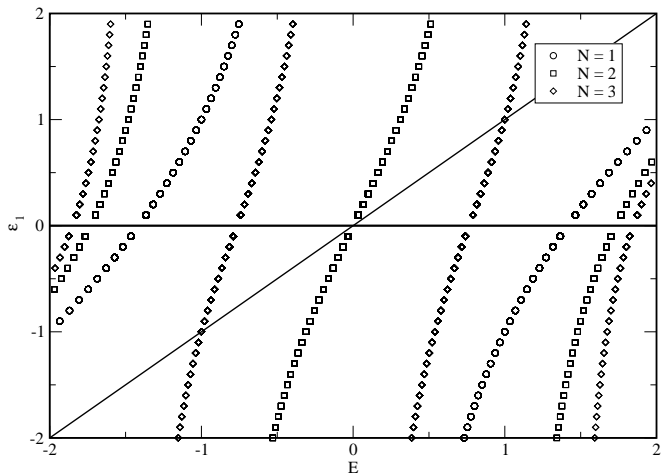}\\
\parbox{8.9cm}{\caption{ \label{fig2}
\footnotesize {Curves satisfying the condition $T=1$ for $E$ and
$\epsilon_1$ values as indicated. Only the cases $N=1,2,3$ are
displayed.}
}}
\end{center}
\end{figure}
Using the transfer matrix (\ref{mtemp}), we can apply equation 
(\ref{trc}) to get the transmission coefficient for this
system. Since the presence of the generalized dimers in the chain
introduce correlations in it, it is expected that
particular values of the energy will produce transparent states.
Unlike the RDM, where any site energy equal to the resonant energy
produces a transparent states, we find that for this system,
only certain values are allowed.

A relevant feature of this system is that, aside for this set of
energies, there are also other values of $E$ for which the transmission
is also equal to one. For any value of the impurity energy,
one can always find one or more resonant energies. The number of 
resonant energies increases as the
number of host sites between the impurities.
This can be seen in Fig. (\ref{fig2}), where we have plotted pairs of
values of energy and site energy that produce a transparent state. Only the 
cases for 1 to 3 host sites are shown. Even though the figure shows
a blank space between the points, the real data is continuous.
For example, the relationship between $\epsilon_1$ and
$E$ is 
\begin{eqnarray}
\epsilon_1=\frac{E^2-2}{E}&\mbox{for}&\qquad N=1,\\
\epsilon_1=\frac{E^3-3E}{E^2-1}&\mbox{for}&\qquad N=2\,\,\mbox{and}\\
\epsilon_1=\frac{E^4-4E^2+2}{E^3-2E}&\mbox{for}&\qquad N=3.
\end{eqnarray}

Similar expressions can be
found for other values of $N$. The horizontal
line is the case of zero site energy (regular lattice) for which we know that
a continuous set of resonant energies exist. The diagonal line across
the figure has been added to show those cases for which $E=\epsilon_1$
(see also table 1). 
From the crossing of this line with the curves we notice 
that the case of $N=3$ has two values of energy (-1 and 1). 
Apart from the trivial case $E=0$, the
$N=1$ and $N=2$ cases have no extended states when $E=\epsilon_1$. Cases
of $N\ge 3$ will always have resonant energies, but only for very
particular values.

\begin{center}
\begin{tabular}{|c|c|}\hline
N & $\epsilon_1$ \\ \hline
1 & 0 \\
2 & 0 \\
3 & 0, $\pm 1$ \\
4 & 0, $\pm\sqrt 2$ \\
5 & 0, $\pm\frac{1+\sqrt 5}{2}$, $\pm\frac{1-\sqrt 5}{2}$  \\
6 & 0, $\pm 1$, $\pm\sqrt 3$ \\
\hline
\end{tabular}\\
\vspace*{0.8cm}

\footnotesize {Table 1. Resonant energies for the case $E=\epsilon_1$. Only
systems with $N$ smaller than 7 are shown.}
\end{center}

In Table 1, the resonant energies when $E=\epsilon_1$
are given for $N=1$ to $N=6$. Other values for $N>6$ can be easily 
found. The important point here is to notice the following two
observations: 1) the number of resonant states is twice the number
of the host sites in the generalized dimer ($2N$). 2) The resonant energy
for $N$ host sites is also a resonant energy for any multiple
of $N$. For example, for $N=3$, we have $E=1$, which is also a resonant
energy for $N=6,9,12...$. Likewise, $E=\sqrt 2$ is a resonant energy
for $N=4,8,12...$, and so on.

Aside for this set of
energies, there are also other values of $E\ne\epsilon_1$
for which the transmission
is also equal to one. In fact, for any value of the impurity energy,
one can always find one or more resonant energies. The number of 
resonant energies increases as the number of host sites 
between the impurities.  It is not exactly equal to $N$ in all cases
because it will depend on the crossing of the line of constant
$\epsilon_1$ with the curves in Fig. (2). As we can see, there is a minimum value
for which we will have $N+1$ states.  This is shown in Fig. (3).
Notice, also from Fig. 3 that, given the Fermi energy of the system, 
one can find a resonant
energy close enough to it for a relatively small $N$.

\begin{figure}[t]
\begin{center}
\includegraphics[width=8.5cm]{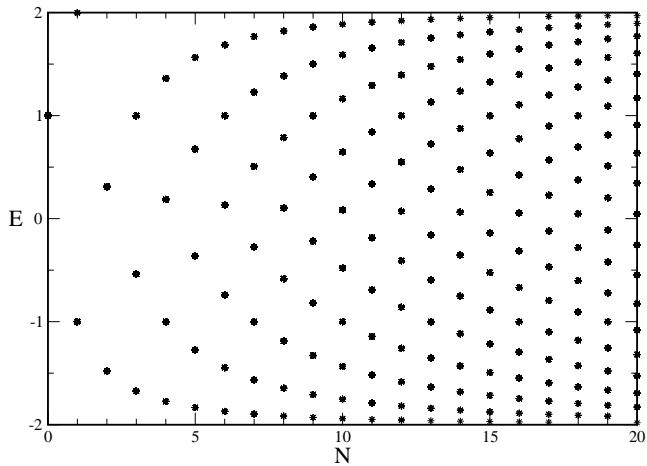}\\
\parbox{9.0cm}{\caption{ \label{cruces}
\footnotesize {Resonant energies as a function of the number of host sites
between the impurities. Here $\epsilon_1=1$.}
}}
\end{center}
\end{figure}

In conclusion, we obtained the formula for the transmission coefficient
in a compact and straightforward way, making use of the Cayley-Hamilton
theorem and the transfer matrix method. We applied this formula
in two cases: 1) for the $n$-dimer system and obtained previous
results and 2) for a new system that we called a generalized dimer,
consisting of two impurities embedded in the host lattice.  This system
was found to have resonant energies for $E=\epsilon_1$ and also for
values of $E\ne\epsilon_1$. The number of these energies grows like
the number of host sites between the impurities, $N$. This is an important
feature of the system, since, by varying $N$, one can find a resonant
energy close enough to the Fermi energy level.
We believe that these kind of systems will be good candidates
for the design of actual physical devices.

\end{document}